# Toward an Ontology for Third Generation Systems Thinking

Anatoly Levenchuk
ailev@asmp.msk.su
Aisystant

## Abstract

Systems thinking is a way of making sense about the world in terms of multilevel, nested, interacting systems, their environment, and the boundaries between the systems and the environment. In this paper we discuss the evolution of systems thinking and discuss what is needed for an ontology of the current generation of systems thinking.

## 1. Introduction

**SYSTEMS THINKING** is a way of making sense of the complexity of the world [1]. Rather than only looking at individual parts, systems thinking takes a more holistic approach in which the world is viewed in terms of multilevel, nested, interacting systems, their environment, and the boundaries between the systems and the environment. Idealistically, the goal is to develop best practices for systems that will change the world for the better. Developing such best practices requires that individuals from many specialized disciplines both understand the systems and communicate with one another in a meaningful and useful manner. The notion of an ontology was developed as a means for supporting such interdisciplinary communication and for helping to better understand subject areas as well as many other purposes. In this paper we survey the generations of systems thinking. We then discuss the requirements for an ontology for the current generation of systems thinking.

We begin in Section 2 with some background in basic ontology and epistemology that will be used for describing systems thinking concepts. Section 3 discusses the first-generation approaches in which a system is described within its environment as it is operating. Section 4 then discusses the second-generation approaches which are concerned with the creation of a system-of-interest by systems that construct and enable other systems. The current generation of systems thinking is then described, starting with continuous development techniques in Section 5. Scaleless system descriptions are covered in Section 6, Section 7 is concerned with systems that perform operations on both physical objects and abstract objects. Section 8 speculates on how the evolution of systems can be explained using thermodynamic principles. The paper concludes by summarizing the most important requirements of an ontology for third generation systems thinking.

## 2. Systems Ontology and Epistemology

An ontology is a way to express what a subject area represents by defining a set of concepts and categories as well as how the concepts and categories are related to each other. An ontology should indicate the objects that would be good to distinguish in the world for reliable active/embodied inference within the subject area [2]. We formulate the task of creating an ontology in terms of attention management, and "explicit formal specification" which here indicates that these objects of attention are not spontaneously singled out but according to some explicit model as ontology specified by using another model as its formalism (such as one of the foundational ontologies).

We resolve the issue of creating an ontology according to Popperian epistemology [3]: objects in an ontology appear by guessing, the acceptability of these guesses for judgments about the world is questioned, but guesses that survive criticism are "taken seriously." Good guesses about how the world is made up of systems (what is the ontology of the system) and judgments that these guesses help change the world for the better can be taken from current literature. Suppose we do not know their critique and cannot immediately suggest falsification of those conjectures by reasoning and/or experiment. In that case, we consider them our best theories (i.e., the state-of-the-art) about the world, take them seriously and then teach people (replicate memes of those theories) to make those conjectures about the important objects shared.

Of course, we do not just tell people the systems ontology. However, we teach them to attend to abstract and physical objects in the world, described by shared explicit and as formal as possible specification, i.e., ontology. Problems emerging within an activity performed with this ontology as a theory of the world mean that ontology is falsified, but this is simply a reason to correct the ontology by correcting identified errors. Systems ontology thereby continuously evolves. It represents at each point in time the best we know about systems at that point. We [4] have implemented a curriculum in which several hundred people a week are now learning to identify in the world objects described/typed by systems ontology. The curriculum implements teaching students fully conscious at the start and then interiorized "automatic" conceptual guidance of students' attention to objects described/typed by systems ontology. The student is trained to:

- to direct one's own and other agents' attention to objects from an explicit specification (ontology), that is, to direct attention conceptually rather than spontaneously (Ontological Modeling course),
- to hold conceptually guided attention on a wide variety of time scales, including collective attention in such agents as a team and whole enterprise via leadership, or influencing on community or society (Self-Collectedness course),
- to direct attention to systems (Systems Thinking course),
- to direct attention to the activities/practices that are carried out by constructor/enabling systems (Methodology course),
- structure the activities and roles that are necessary to be held in attention in the course of systems engineering projects (Systems Engineering course),
- to draw the attention of oneself and the team to the objects described by the systems ontology specialization for systems-constructors such as organizations/businesses (Systems Management course).

To teach students a state-of-the-art systems ontology, we have harmonized several particular conceptualizations of the concept of systems (i.e., we performed a merge of partial systems ontologies), which after merging, represent together an ontology of the third generation of systems thinking. As usual, each generation of systems thinking incorporates all the previous generation's achievements but adds something new.

## 3. The first-generation approach

The first generation of systems thinking emerged in the 1940s, mainly as a result of the work of von Bertalanffy [5]. The notion of a system as a subject separated from its environment appeared in physics long ago. However, the systems approach as a consideration of the whole world as interacting systems appeared mainly after von Bertalanffy's works on general systems theory. "Approach" is the usual term for a situation in which ontology, successfully developed and tested in one domain, begins to be used in many different domains. The key here was the realization that systems are interacting holons (Koestler coined the term [6] to describe a part of the whole that itself is made up of parts). Also, these systems somehow appear in the world and then disappear from the world during a life cycle (von Bertalanffy was a biologist, and he generalized, among other things borrowed from the systems approach in biology, that the life cycle is the birth-growth-breeding cycle). The interaction of parts of the system during work/operations results in emergent properties of the whole. The gears in the clock do not yet show the time, the clock shows the time, and the house with the clock inside no longer shows the time. The key here was that the first generation of systems thinking operated on at least two different ways of partitioning depending on the time of consideration:

- functional partitioning depends on the system's purpose in the suprasystem and the subsystems in the system. This viewpoint needed to think about the system during its operations in its environment.

- a constructive/modular partitioning is needed to think about the system at the creation/construction time.

The difficulty in mastering systems thinking was that people needed help grasping the concept of selecting/isolating dynamic objects in the world with their attention to the system at operations time. Often, they imagine an "explosion diagram" in their heads when mentioning decomposition into functional parts, simultaneously losing the multiple system levels of such a partitioning [7]. Mereology is the study of parts and the wholes that they form. One cannot discuss the emergent properties of the stopped broken-down system into single-level physical constructive parts because interaction happens in operations time, and viewpoint must be functional, not constructional. Nevertheless, partitioning into constructive parts is also essential, for such a system must be created in construction (design, implementation, test, deployment, transfer to operations) time. For example, scissors functionally consist of a cutting unit and a handle, while structurally, for construction, they consist of two halves of scissors and a screw that fastens them together. The scissors user is concerned with the functional viewpoint, while the factory engineer is concerned with the constructional viewpoint.

Bertalanffy included systems engineering in his proposed set of systems disciplines. The field of systems engineering subsequently developed rapidly. In systems engineering, the target systems were physical, which has the advantage that all descriptions have a guaranteed grounding. However, it proved to be impossible to go beyond mereology to more abstract notions such as the relations between whole objects, parts, parts of parts and boundaries between parts. The study of such relations is called mereotopology. Attempts to develop a mereotopology of mental/abstract/mathematical objects that comprise views mostly failed.

## 4. The second-generation approach

The second generation of the systems approach relied mainly on the ideas of systems engineering: physical (including cyber-physical, including cyber-physical with people) systems-of-interest are created by people with tools, not reproduced themselves in a cycle, and usually non-living. The notion of the life cycle turned out to be neither life nor cycle, although the term for the concept remained in use. In the second generation of the systems approach, constructor/enabling systems appeared. That introduces, besides the part-whole relation, the construction/enabling relation and the viewpoints for the activity of constructor systems and interaction of constructors and systems-of-interest (the most famous introduction of this viewpoint was works of Checkland in the late 70s and early 80s [8]). Systems engineering was supplemented with systems engineering management from the viewpoint of construction/enabling systems, thus enabling a chain of constructors and enablers (i.e., constructors of constructors of systems-of-interest).

The systems engineering diagrams emphasized the "waterfall" model of the life cycle, in which the system was conceived, designed, implemented, tested, operated, and decommissioned, where the project usually ended. The living system does all this work for itself by itself (with some conceiving and design work done for them by evolution, but this was not usually considered). However, in engineering, it was all done by the constructor as a separate system. Different roles (stakeholders) in the project-as-the-construction-chain have different interests/concerns for the system and the project. These concerns require different kinds of descriptions/views that should be made with respective viewpoints as methods of system description that suit to reconcile the very different concerns of the very different project roles.

The ISO 42010 [9] and ISO 15288 [10] standards solidified the conceptualization of the system for systems engineering. Wide usage of these standards ensured that the second-generation systems engineering ontology of the systems approach was shared. A formalization of systems ontology based on 4D extensionalism (the idea that if two objects occupy the same place in space-time, they are the same object – and it is a physical object) was proposed in ISO 15926-2 [11] in 2003. It drew on ideas from the Business Objects Reference Ontology [12]. Based on this 4D ontological commitment, several other similar ontologies were proposed mainly for military applications. This work was led by the IDEAS Group [13]. Special mention can be made of the work of Matthew West, who proposed the High Quality Data Models ontology based on the same ideas of 4D extensionalism [14], in which the concept of the system was made one of the central. All these approaches assumed not just conceptualization specification but also a formal expression of conceptualization. This formalization can be done with logical ontology description languages (such as the very rarely used EXPRESS language [15] but also appeared later more popular Web Ontology Language (OWL) [16]) to create data models in Product Lifecycle Management (PLM) systems databases [17]. The fundamental ontological premise was to use a 4D ontology, which gave a reasonable and compact description of the system's changes during its construction from parts [18].

The peak of this line of work on the second generation of the systems approach that evolved in systems engineering came in 2008-2013. After that moment, public interest in such logic-based formal systems ontology descriptions faded a bit. The development of ontologies as an explicit formal specification of shared conceptualizations was no longer seen as an advance in AI creation, and not even the rebranding of formal ontologies as knowledge graphs helped [19]. The semantic web was expected as an internet mainstream Web 3.0, but this idea was never realized. Semantic technologies remain niche.

## 5. Continuous development

The "waterfall model" [20] was the limited idea of a single unidirectional passage of the life cycle as a set of works carried out by constructors following life cycle practices. System-of-interest was considered as passively undergoing this one-time creation. Therefore, agile approaches were developed in engineering, overcoming the concept of a "waterfall" life cycle. In 2001 the Manifesto for Agile Software Development appeared [21]. In 2017, evolvability as a core architectural characteristic reflecting the "continuous everything" notion was entrenched in best engineering practice and entrenched in the literature [22]. The architecture was finally separated from development into a separate domain as the literature began to reflect the inevitable "productive conflicts" between architects and developers. Architecture worked on slicing the system into minimally interacting modules to maintain the stability of the whole in the evolutionary time scale (operations characteristics such as -ilities well known in engineering appear as architectural characteristics). At the same time, developers were concerned with maximizing functionality, which could entail temporary deterioration of architectural characteristics as an accumulation of "technical debt" (work aimed at maintaining selected architectural decisions but not directly affecting functionality).

The idea of evolutionary cycles was introduced to engineering, but in a different way from biological evolution because the systems memome, as an analog of the biological genome, was not contained in the manufactured system but was separately stored in a digital form somewhere in the constructor systems. Digital form implies exact multiple replications without accumulation of errors of analog representation. The separation of the memome and the system defined by that memome significantly speeds up the techno-evolution compared to the biological one. For some techno-crab, it was not necessary to wait until the whole crab died to replace the unsuccessful claw variant only with the crab as a whole by mutations of genes forming the phenome of the claw. It was also possible to use smart mutations to techno-crab memome, getting quicker and more reliable results. It is possible to try only claw variants that showed success in simulations in the virtual world – and to change in real life only the claw, but not the whole crab, especially if the system has a proper architecture with good modularity. Techno-evolution gives results unattainable only by a one-time offer of a set of new features (one-time "waterfall" engineering) or only evolution with random mutations [23]. System modeling at operation time practice was adopted with a digital twin to assess the system's success in its operations. The concept of a digital twin reflects not only the memome but also the phenome of the system. So a systems ontology focused on two times (1. operation and 2. construction of a single increment/feature) was not enough. It was also insufficient to model a whole group of systems (product lines [24], sometimes system families) as different system variants coexisting.

The way out was the idea of "continuous everything" [25] as infinitely ongoing system development with continuous delivering and commissioning of more and more variants of the system-of-interest with more and more changes for fit to environment. Slowing down development with the growth of system-of-interest complexity in continuous delivery is suppressed both by the evolving architecture and the changing structure of the teams that perform this development (using Conway's inverse maneuver [26]). In software engineering, this idea appeared within the DevOps/SRE/platform engineering approach [27]. In traditional systems engineering, this idea is discussed now in examples of aerospace systems. Testing of the next version engines nowadays starts before the tested engines of the first version in the rocket perform their first flight, and each new instance of rocket hardware has an improved design. "Continuous everything" means moving engineering to techno-evolution projects, that is, evolutionary changes in the memome – continuously improving the information model of the system design, not just the improving phenome by "field refinement" of an instance of the system. Continuous development as the practice of techno-evolution tries to endlessly keep system-of-interest, adapting it as long as possible to changes in the environment, in itself, and changes in a chain of constructor systems. In biology and life sciences, similar problems of approaching infinite in time development/techno-evolution are discussed as open-endedness [28].

Thus a requirement to reflect the ideas of evolution in systems ontology emerged. Suppose there was a developed ontology of evolution. In that case, it could be used to develop a third-generation systems approach that would account not only for construction/enabling systems as agents in their different project/activity roles but also for techno-evolution. It would work explicitly with three times:

1. Operations time of the instance of a system with intended by design phenome
2. Construction time of a system instance with an increment change in design, i.e., phenome from memome implementation
3. Continuous changing of memome time (techno-evolution, continuous everything)

## 6. Scaleless descriptions of physical systems

In the discussion of evolution in biology, the systems approach manifested itself in reality in major evolutionary transitions: increasing complexity in systems from large molecules to cells, from cells to multicellular organisms, and from organisms to populations. Approximately the same increase in the complexity of systems occurs in techno-evolution: chips are composed of transistors, cards of chips combine to form computers, stacks of computers aggregate to form data centers, and data centers transit to global computer networks. The analogy seems straightforward but requires a scaleless theory that will be common for systems of any scale and origin. Such a theory should be valid for the world of elementary particles in nuclear physics (taking into account quantum phenomena) but also for macro-objects, including living and non-living, as well as living conscious beings and collections of living conscious beings (for example, humankind as a whole). Scaleless also meant scaleless in the fourth dimension, time: considering the system construction/growth time, operations time, and evolution time.

A key step for developing scaleless descriptions is to formulate physical phenomena in informational terms and to explain how some objects remain stable in the physical world. How do some objects (such as a molecule or a person) maintain their shape in space-time? Fields, Glazebrook, Levin, and Friston proposed an ontological framework of panpsychism in a variant of minimal physicalism to describe physically persistent systems as the realization of the principle of free energy minimization [29]. Their theory covers the complexity spectrum from elementary particles to people and society. Free energy is defined as an informational characteristic of a system rather than the traditional energy of mechanical work or the work of electromagnetic forces. These systems increase in complexity from elementary particles through molecules, through inert matter bodies to living cells, and from unicellular to multicellular organisms and their populations. All of these types of systems reduce the Bayesian (or excess Bayesian to account for the quantum-like nature of computation in biology [30]) surprise of mismatch between expected measurements according to a generative model of the world and actual measurements in the world.

This scaleless ontology for physically persistent systems has been mereologically formalized using category theory as a foundation ontology [31, 52] and expressed in a form that allows describing quantum-like active/embodied inference [32]. This line of ontological engineering provides concepts for accurate reasoning about systems of very different evolutionary complexity, including applying stochastic methods to non-ergodic systems, that is, systems with memory. One of the strongest confirmations of such a theory was the creation of a learnable hybrot [33]. Hybrot, as a sufficiently complex system capable of learning (such as a natural neural network), must exhibit (due to the principle of free energy minimization) behavior that minimizes the unpredictability of the external environment. Precisely that was shown by experiment, hybrot learned to play Pong [34]. This line of research achieved the formalization and mathematization of ideas of the first generation of the system approach in physics-based ontology. Due to the scaleless nature of this ontology, other time scales also can be covered: the time scale of a system's life cycle and the time of evolution. The key to this was the notion of measurement, developed in quantum physics to solve the problem of the observer. Measurement is an interaction of systems, not just passive perception/observation. The reverse is also true: the interaction of systems in the construction process is described as the inverse of measurement. Any molecule's interaction as a system with its environment can be considered either a measurement or construction. A molecule is a proto-agent that somehow perceives/cognizes/measures the world around it, maintaining its stability while obeying the principle of minimizing free energy.

This line of reasoning (perception/measurement as inverse to changing/constructing, and both are an interaction of systems) was drawn in the constructor theory proposed by Deutsch and developed by various researchers of "quantum gravity" that needed scaleless theories [35]. A constructor is a physical device that can maintain its immutability for a long time while repeatably changing its environment by some pre-described sequences of operations. Examples can be a catalyst molecule, a robot with a universal computer, or a living being. Sufficiently advanced constructors can replicate themselves, among other things, to be the subject of evolution. Scaleless physical theories are based on notions of information-related changes as computations have proven to be very productive.

## 7. Constructivism

The next thing to be considered in the physics-based system ontology is the ontology of physics and mathematics itself as a foundation ontology. Such an ontology has been proposed in many works, but let us highlight the work of Deutsch [36]. He proposed to consider physics as a study of physical (situated in space and time) objects, mathematics as a study of mental/abstract/mathematical objects, and computer science as the natural/experimental science of proving that the behavior of physical objects can somehow reflect the behavior of ideal/abstract/mathematical objects. Computer science, according to Deutsch, is the science of universal computers as physical devices capable of performing computation – traditional electronic, quantum, and many other types, including biological and social computers such as the human brain and even collectives of humans altogether with their computers.

These ideas imply a shift of the discussion from static models and passive data to universal computers as physical devices that interpret such data and change the state of the environment depending on these computations. The input of raw data for computations and output of results in the symbolic form is just a particular case here. Perception of the environment and changing the environment and/or "self" as a computer/constructor device/system is just a more general case. Central to this approach will be the notion of a constructor. This physical device can interact with the environment, repeatedly performing some operations upon the objects in the environment while maintaining its stability (e.g., a catalyst molecule, a robot, or a human). So, we have some approach to a formal description of systems of the second generation of the systems approach: some systems of different degrees of agency (from inert matter, robots, intelligent agents, and all kinds of their hybrids) construct according to some design and method data, different types of systems-of-interest, which further operates in their environment.

One more attempt at integrating physics, topology, logic, and computer science with a shift of attention to enactive operations/morphisms with objects instead of considering static objects in their static relations was proposed by Baez and Stay in the Rosetta stone approach [37]. After moving from describing system interactions as processes in networks, Baez proposes to use the formalism of symmetric monoidal category theory to describe not t just t processes but t open systems interacting with the environment [38]. Networks such as electrical circuits, hydraulic networks, and network interactions in system dynamics are usually associated with functional representations of the systems, that is, representations of the system in its environment in operations time, the first generation of the systems approach. Again we see a move to use category theory as a foundation ontology for systems ontology to represent it not as a set of static objects and relations but as morphisms as object changes. This move corresponds to a general move toward constructivism in mathematics, where we change eternal classes and their relations to construction operations [39]. This move makes it possible to reformulate mereology as the central ontological discipline of the systems approach from the study of the objects in "part-whole" relations [40] to the study of the operations of constructing a whole from parts.

Another move in this direction of constructivist mereology is Fine's ideas about the mereology that includes abstract objects [41], which suggest operations of constructing the whole from parts. Operations of constructing the whole from parts also apply to descriptions (constructing a set from its elements). Although Fine does not explicitly say so, defining construction operations (as well as category theory morphisms) performed as if by "nobody", but behind these constructor operations in the physical world, one can easily see a physical device that is the constructor from constructor theory. For the constructor of the whole for abstract parts, there will be a physical device implementing a "universal computer". It is also a constructor from constructor theory, including quantum computers and living mathematicians – they are computationally equivalent and physically embodied to get information about input and output data of computation. Fine's approach allows us to extend the notion of a system beyond physically interacting parts. Interaction of abstract objects appears inside computer devices as physical parts of computers interact to emulate abstract objects. The definition of a system as interacting parts with emergence persists, but interaction appears in the constructor/computer system rather than in the abstract system-of-interest. It is possible to discuss systemicity in complex communities in the physical world despite the absence of the interaction of the members of these communities with each other. The interaction of community members occurs by carrying out operations of reasoning about these communities in the constructor's computer (e.g., social engineer's brain) [42].

Core Constructional Ontology [43] is based on ideas of Fine and offers a constructivist perspective on the theory of parts, sets, and relations. Core Constructional Ontology serves as the second layer of foundational ontology for systems descriptions in a PLM system of engineering projects. For core constructional ontology first layer of foundational ontology is mathematics. 4-Dimensionalist Top Level Ontology expressing 4D mereotopology of space-time gives us the next ontology layer as an upper ontology for modeling physical systems in current systems engineering [44].

Using some mathematics as the foundational ontology is typical for computational ontology in the form of variants of first-order logic. However, more and more often, foundational ontology is based on morphisms and, respectively, category theory. Using mathematics as the foundational ontology and omitting all other (upper ontology, middle ontology) ontological levels are typical in physics. Physics itself, in terms of systems ontology, deals with functional objects (e.g., "physical body") whose roles are played by different objects of the physical world (mountains, stones, molecules, elementary particles). Then in conjectures and experiments, the peculiarities of behavior of these functional objects are found, and mathematical relations express them. Structural similarities of formulas describing some different objects raise the question about the general ontological nature of these objects. Thus ontology merges can be performed by analogical reasoning with ontologies represented by mathematical formulas. This technique is also used for ontologies expressed logically as lattices (knowledge graphs), an example of which is the VivoMind Analogy Engine [45]. The usage of such a technique by physicists is ubiquitous. E.g., in works on information theory, "free energy" is called so due to its appearance at the same place in similar formulas in informatics theory that it appears in formulas of thermodynamics [46]. The trend is to use mathematical/abstract objects and formulaic relations between them to formulate a third-generation systems ontology. Then we get the physical explanation of evolution with imminent major evolutionary transitions leading to multiple system levels in persistent objects.

## 8. Thermodynamics of system evolution

Vanchurin, Wolf, Koonin, and Katsnelson observed that analogy reasoning with classical thermodynamics and information theory, quantities in thermodynamics, machine learning, and evolutionary biology lead to a surprising common framework/ontology for all of it [47]. In other words, evolution could be described as learning, and learning has thermodynamic nature, i.e., physical phenomenon. Based on these notes, they formulated the theory of evolution as multilevel learning [48]. In this theory, the driving force behind evolution is frustrations arising from conflicts between objects of different system levels. This notion of frustrations was introduced into the systems language by studies of spin glasses as an example of the behavior of non-ergodic (with memory) systems. It meant frustrations analogous to geometric frustrations as the impossibility of stable geometry of spins in glasses [49]. Evolution is thus a learning process, which is reduced to solving the optimization problem of finding a minimum of the free energy by changing the structure of numerous system layers. In the course of this, evolution finds quasi-minima but not an absolute minimum. From time to time, there is a jump in complexity growth (appearance of another system level, another type of whole system, e.g., multi-cell organism constructed from cells), which gives a sharp minimization of the free energy of the evolving system. However, this is still just another quasi-minimum, not an absolute minimum. In [50], the physical/thermodynamical nature of the growing complexity of systems was shown with the example of biological systems. The reasoning there is scaleless, i.e., applicable to the living (with different degrees of consciousness and sentience), cyber-physical, and even inert systems. It means that systems ontology should include the notion of conflicts between system levels, the notion of the frustrations caused by these conflicts, and the rule (law, conclusion) of the inevitable growth of system complexity (inevitable increase in the number of system levels in the course of evolution).

Many of the aforementioned researchers are in contact with each other. Therefore, there is hope for a shared third-generation system ontology. For example, there were discussions between Vanchurin and Friston, which is indicated at least in the acknowledgment to [47]. Most of the mentioned works are based on thermodynamics and the premise that all systems obey the free energy principle during their existence, creation/construction, and subsequent continuous evolution. This systems ontology is currently expressed in the formulas familiar to physicists, traditional for thermodynamic calculations. The translation of the traditional formulas for expressing this ontology into the constructivist form of category theory is a separate topic. However, such a formulation of the problem is more or less usual for mathematicians. We can discuss the research program on unifying mathematical languages used in different studies of the same domain of thermodynamic-based systems ontology. The same can be said about reformulating the differential form into a quantum-like form to increase the speed and accuracy of the physical modeling of biological evolution and techno-evolution [51].

## 9. Conclusion

An ontology for state-of-the-art third generation systems thinking should:

- provide types of objects for layered directing of attention to ensure the evolution of systems-of-interest (continuous everything) by constructor systems

- consider at least three times of system existence: operations, construction and evolution of phenome, evolution and development of genome and memome.

- be based on physics, mathematics, and computer science

- treat systems as stable entities within a minimalist physicalism (including systems active to themselves and their environment, systems seeking minimum free energy by active/embodied inference, regardless of their level of "intelligence")

- give scaleless descriptions of systems (phenomena of quantum physics are thereby taken into account), which explains the emergence of system levels (growth of complexity) due to multilevel optimization to achieve the minimum of free energy

- no longer express mereology through eternal classes and relations between them, but rather through morphisms and operations reflecting operations with physical systems in the course of their interaction as well as operations with abstract objects performed by constructor systems with a (universal in the sense of Turing machine equivalence) computers in them.

**ANATOLY LEVENCHUK** has been a strategy consultant for more than 20 years. He helps with vision and strategy definition for many government agencies and large private holding companies in Information Technology and Internet businesses, the securities market, the electric power and transmission industry, the nuclear industry, the shipbuilding industry and transportation infrastructure. He is currently the scientific director of Aisystant.